\def   \ni {\noindent}

\def   \ssk {\vskip  5truept}

\def   \newline {\hfil\break}

\documentstyle[epsfig]{article}
\begin{document}

\hsize 5truein
\vsize 8truein
\font\abstract=cmr8
\font\keywords=cmr8
\font\caption=cmr8
\font\references=cmr8
\font\text=cmr10
\font\affiliation=cmssi10
\font\author=cmss10
\font\mc=cmss8
\font\title=cmssbx10 scaled\magstep2
\font\alcit=cmti7 scaled\magstephalf
\font\alcin=cmr6 
\font\ita=cmti8
\font\mma=cmr8
\def\ref{\par\noindent\hangindent 15pt}
\null


\title{\ni BeppoSAX observations of bright Seyfert 2 galaxies
}                                               

\ssk \ssk
\author{\ni L. Bassani, M. Cappi, G. Malaguti\\
On behalf of the Compton Thin Seyfert2s collaboration}  
\ssk
\affiliation{ Istituto TeSRE/CNR, Via Gobetti 101, 40129 Bologna, Italy
}                                                
\ssk
\baselineskip = 12pt

\abstract{ABSTRACT \ni
We report on BeppoSAX broad band (0.1-100 keV) observations of bright 
Seyfert 2 galaxies and
discuss the results in the framework of the unified theory of AGNs. 
The  data are used to probe N$_{\rm H}$ distribution 
in type 2 objects and 
to identify heavily absorbed (10$^{24}$ $\le$ N$_{\rm H}$ $\le$
10$^{25}$ cm$^{-2}$) sources, which are particularly bright in hard X-rays.
They  also provide, for the first time, high energy spectral data
on Seyfert 2s and tight constraints on the shape of their intrinsic 
power law continuum. These results 
confirm the basic expectations of the unified theory and  indicate
where the INTEGRAL mission can give a substantial contribution.
}                                                    
\ssk
\baselineskip = 10pt

\ssk
\baselineskip = 12pt


\text{\ni 1. INTRODUCTION
\ssk
\ni

Many attempts have been made in the last decade to explain the 
complex phenomenology observed in Active Galactic Nuclei (AGN) within a 
common 
framework, i.e the AGN unified model. For Seyfert (Sey) galaxies, 
the main discriminating parameter
is the inclination of our line of sight with respect to an obscuring torus
surrounding the central source (Antonucci 1993); accordingly one expects
the direct continuum of type 1 and 2 objects to be the same except
for the obscuration by the torus material. 
X-ray observations are probably the best observational 
tool to test this unifying model both directly by measuring the torus absorption
in the source 
and indirectly by studying its reprocessing effects on the intrinsic 
spectral shape; once these torus related components are 
properly removed, the shape of the primary continuum can be determined and  
compared to the one observed in  Sey1s (see Matt et al. 1999).
\ssk
\ni 2. PROBING ABSORPTION IN SEY2s 
\ssk 
\ni
Table 1 lists for a small sample of Sey2s, the spectral parameters
deduced by fitting their broad band X-ray spectra with a baseline model
which includes an absorbed power law, a reflection component due to the torus
(pexrav model in XSPEC with a cut-off energy frozen at 1 MeV) and a narrow 
iron K$\alpha$ line (errors are at 90$\%$ confidence level).
The Table also lists the parameters T 
(F$_{\rm 2-10 keV}$/F$_{\rm [OIII]}$) and H 
(F$_{\rm 20-100keV}$/ F$_{\rm 2-10keV}$). 
One striking result of Table 1 
is the wide range of absorption found:  although the sample is small,
the column density values span 3 orders of magnitudes and 4 out of 5 objects 
have N$_{\rm H}$ $\ge$ 10$^{23}$ cm$^{-2}$; accordingly the T ranges from
Sey1s values (Maiolino et al. 1998) to values $\ll$ 1. Sources at high 
N$_{\rm H}$ tend to have  flat 2-10 keV spectra (when fitted with just a 
simple absorbed power law model, i.e.$\Gamma$ $\sim$ 0.4 for Mkn3, 
Cappi et al. 1999 and 1.1 for NGC7674, Malaguti et\\
al. 1999)
 and a
strong iron  K$\alpha$ line (EW $\sim$1 keV): all together these values 
indicate that the 
X-ray primary emission is heavily depressed although the continuum,
reflected by
the torus inner surface, is still visible.
This is in line with
recent BeppoSAX results on weak Sey2s, which
indicate that the average  N$_{\rm H}$  of type 2 AGNs\\
\\
\\

is very likely much higher than deduced in previous surveys
(Maiolino et al. 1998). This is expected in the unified theory as Sey2s
are viewed through the torus, which 
is expected to be very thick and compact (Pier and Krolik 1993).
Bassani et al. (1998) have 
demonstrated that indeed 25-30$\%$ of Sey2s have N$_{\rm H}$ $\ge$
10$^{24}$ cm$^{-2}$. At these column densities, the 2-10 keV spectrum is
Compton thick and the source is most likely reflection dominated.
However, for N$_{\rm H}$ in the range 10$^{24}$-10$^{25}$ cm$^{-2}$, 
the source
is still "thin" to Compton scattering as X-rays
in the 10-100 keV band can still penetrate the torus allowing an
estimate of the source column density. These borderline objects turn out to
be particularly bright in hard X-rays (above 10 keV), as at these energies 
the transparency effect is coupled with a strong reflection. Only a few of 
these  sources are presently known: Mkn3 (Cappi et al.
1998), NGC4945 (Done et al. 1996) and possibly NGC6240 (Turner et al.
1997). More recently, BeppoSAX discovered  two  more sources: 
Circinus (Matt et al. 1998) and IRAS09104+4109 (Franceschini
et al. 1998). Mkn 3 can be taken as a prototype borderline object (see 
Table 1): the X-ray flux is highly absorbed but in the hard X-ray band
the source has a flux similar to that of 3C273 (one of the brightest 
objects of the high energy sky) and so a hardness ratio
H$\gg$10 (see Figure 1).
Similarly in NGC4945 and Circinus, H is
48 and 13 respectively (Done et al. 1996, Matt et al. 1998);
these values suggest that also NGC7674 may have N$_{\rm H}$ 
in the range 10$^{24}$-10$^{25}$ cm$^{-2}$, although a higher absorption 
cannot be excluded.
These sources may be particularly
relevant for INTEGRAL observations as they may be likely
candidates for serendipitous source identifications. In particular,
INTEGRAL can assess their incidence in the local universe and
thus better define the N$_{\rm H}$ distribution; 
this is important both to constrain the properties of the
obscuring torus (and hence the validity of the unified model) and to assess
the AGN contribution to the X-ray background.
\begin{center}
{\bf Table 1: Bright Seyfert 2 Galaxies sample}\\
\begin{tabular}{ l  c  c c c c c }
\hline
Source & N$_{\rm H}$$^{*}$ &
EW$_{\rm K\alpha}$ $^{\dagger}$   &R&
$\Gamma$&
T & H  \\
\hline
NGC2110&4.1${\pm0.2}$ & 180${\pm30}$ & $<$0.2 &1.7${\pm0.1}$&9.4 & 1.9\\

NGC7172 &11.5$^{+12.5}_{-10.7}$&224$^{+650}_{-54}$&
1.0$^{+2.7}_{-0.4}$  &1.90${\pm0.3}$    &250 & 3.3  \\

NGC4507     &70$^{+15}_{-15}$           &120$^{+292}_{-68}$
&0.4$^{+0.2}_{-0.2}$  &2.0${\pm0.1}$    &22.8 & 4.2  \\

Mkn3        &130$^{+15}_{-25}$        &997$^{+300}_{-307}$
&0.9${\pm0.1}$  & 1.8${\pm0.1}$   &0.1 &17.4   \\

NGC7674     &$>$1000     &900$^{+470}_{-299}$
&$>$ 40        &1.9${\pm0.2}$     &0.3 & 20.0   \\
 \hline
\end{tabular}
\\
\end{center}
$^{*}$ in units of 10$^{22}$ cm$^{-2}$,
$^{\dagger}$ in units of eV,
\ssk
\ni 2. TESTING THE PRESENCE OF A SEY1 NUCLEUS IN TYPE 2 OBJECTS
\ssk
\ni
A crucial test for the unified theory is the detection in every Sey2s
of the hidden Sey1 nucleus, i.e. of the steep power law continuum 
($\Gamma$ $\sim$ 1.9) breaking at high energies ($\ge$ 150-200 keV, Matt et al.
1999). Measuring the intrinsic spectrum of Sey2s is not straightforward
since the effects of absorption and reflection are strong. For examples,
most of the sources listed in Table 1, besides being strongly absorbed,
have also a reflection component parameterized by R, i.e. the relative amount 
of reflection compared to the
directly viewed primary power law. However, 
once these effects are properly considered, it is 
possible to estimate the intrinsic power law photon index, which in our cases
is well constrained in the range 1.7-1.9.
This is also true for those sources previously reported as  flat,
i.e NGC2110 and NGC7172 (Smith and Done 1996, Guainazzi et al. 1998).
The case of NGC2110 is particularly interesting: in this source,
the 2-10 keV spectrum has always been characterized by a flat power law
index of 1.5-1.6 also observed by BeppoSAX; however the spectrum steepens
to 1.9 when only data above 15 keV are considered. This is clearly shown
in Figure 2, where the broad band data are fitted with a double 
power law with  a break  at 10 keV: low
and high energy data are different at $\gg$ 90 $\%$ confidence and 
$\Gamma_{PDS}$ is consistent
with the canonical Sey1 spectrum. Thus the flat spectrum observed 
below 10 keV is most likely due to a complex absorption
(Malaguti et al. 1999).
Furthermore, the average Sey2 spectrum (20-200 keV) obtained by summing
all PDS data (except NGC7674),  gives an index of 1.8 $\pm$ 0.05
and no deviation from a power law up to $\sim$ 200 keV (Bassani et al. 1998). 
We have  also tested Mkn3 spectrum for a 
high energy cut-off: when this is left free to vary in the baseline
model discussed above, the cut-off energy is found to be   
above 150 keV, independently of the value of R 
(Cappi et al. 1999). 
We can, therefore 
conclude that the primary spectrum bears
no differences from what observed in Sey1s (Matt et al 1999).
Higher energy data, such as those INTEGRAL will provide, are 
necessary to better constrain this primary emission and further compare it
with that observed in Sey1s; obviously the main objective in this field
would be to measure the high energy cutoff of a sample of objects
similar to that used in the present work.

\begin{figure}
\centerline{\psfig{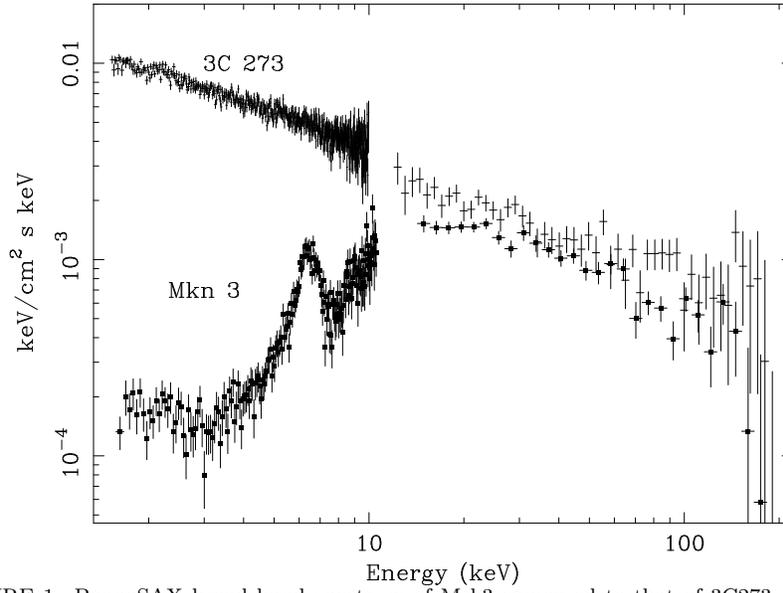}}
\caption{FIGURE 1. BeppoSAX broad band spectrum of Mrk3 compared to
that of 3C273 to show that the emission is absorbed below 10 keV, but
similar to the bright QSO above this energy. 
}
\end{figure}

\begin{figure}
\centerline{\psfig{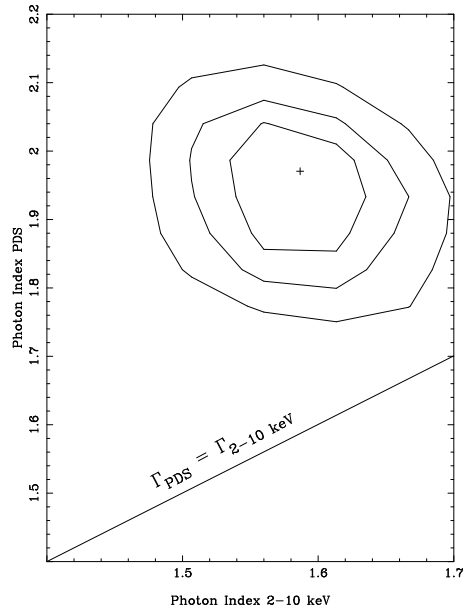}}
\caption{FIGURE 2. NGC2110 photon index in the 2-10 keV band compared to 
that in the 20-100 keV band to show the steepening of the spectrum 
at high energies
}
\end{figure}

\ssk
\baselineskip = 12pt
{\abstract \ni ACKNOWLEDGMENTS
We thank our collegues in the collaboration (P. Blanco, A. Comastri, M. Dadina, G. Di Cocco, 
D. dal Fiume, A. Fabian, F. Frontera, G. Ghisellini, P. Grandi, 
M. Guainazzi, F. Haardt, 
T. Maccacaro, R. Maiolino, G. Matt, G.G.C. Palumbo, L. Piro, A. Santangelo, 
M. Trifoglio, N. Zhang) for their help and cooperation.
}

\ssk
\baselineskip = 12pt


{\references \ni REFERENCES
\ssk

\ref Antonucci R., 1993, A.R.A.  $\&$ A. 31, 473
\ref Bassani L. et al., 1998, Ap.J. Suppl. in press
\ref Cappi M. et al., 1999, A. $\&$ A. in press
\ref Done C., Madejski G.M.$\&$ Smith D.A., 1996 Ap.J. 463, L63
\ref Franceschini A. et al., 1998, in preparation
\ref Maiolino R. et al. 1998, A. $\&$ A. in press 
\ref Malaguti G. et al. 1998,  A. $\&$ A. 331, 519
\ref Malaguti G. et al. 1999, A. $\&$ A. in press
\ref Matt G. et al. 1999, these Proceedings
\ref Matt G. et al. 1998, Cospar Proceedings, in press 
\ref Pier E.A.  $\&$ Krolik J.H., 1993, Ap.J. 418, 673
\ref Turner T.J. et al., 1997, Ap.J.S. 113, 23
\ref Smith D.A. $\&$ Done, C. MNRAS 280,353
\ref Guainazzi M. et al., 1998, MNRAS 298, 834 
}                      

\end{document}